# Historical Aspects of Post-1850 Cosmology

Helge Kragh*[§]

* *Department of Physics and Astronomy, Aarhus University, Ny Munkegade, 8000 Aarhus, Denmark.*
[§] Lectures at XVIII Special Courses at Observatorio Nacional, Rio de Janeiro, Brazil, October 2013. AIP Proceedings (in press).

**Abstract.** Cosmology as an exact physical science is of new date, but it has long roots in the past. This essay is concerned with four important themes in the history of cosmological thought which, if taken together, offer a fairly comprehensive account of the some of the key developments that have led to the modern understanding of the universe. Apart from the first section, dealing with early views of curved space, it focuses on mainstream cosmology from the expanding universe about 1930 to the emergence of the standard big bang model in the 1960s. This development includes theories we would not today consider "mainstream," such as the steady state model of the universe. The last section outlines what might be called the prehistory of the concept of dark energy, that is, ideas that were discussed before dark energy was actually inferred from supernovae observations in the late 1990s.

**PACS:** 04; 95.

## 1. INTRODUCTION

Attempts to understand the universe in terms of mathematics and natural philosophy go back the ancients Greeks and are covered in a rich historical literature. On the other hand, the developments in the twentieth century that transformed cosmology into a proper physical science have attracted relatively little interest among historians of science. Although there are a few solid and comprehensive works on the subject [1; 2; 34], these and the more specialized scholarly literature are not well known among modern physicists and astronomers active in cosmological research.

This essay reviews some of the important themes in the cosmological tradition that started in the early twentieth century. Einstein's original model relied crucially on the notion of curved space, a concept that can be found much earlier and is the subject of Section 2. The following section describes the foundational phase of modern cosmology, from Einstein's static model over the recognition that the universe is expanding and to the early speculations that it might have had a beginning in time. Much of the development after World War II was related to the controversy between two radically different pictures of the universe, the relativistic evolution picture and the steady state picture. This is the subject of Section 4. The final section deals with the history of two concepts that for a long time developed separately, the quantum vacuum and the cosmological constant. They eventually coalesced into the modern idea of dark energy.

## 2. CURVED SPACE BEFORE AND AFTER EINSTEIN

### 2.1. Non-Euclidean Geometries

Whereas curved space as a mathematical concept dates from the early nineteenth century, it took most of a century before it permeated to the physical and astronomical sciences [3]. The eminent mathematician and polymath Karl Friedrich Gauss arrived as early as 1816 at the conclusion that the ordinary, flat or Euclidean geometry was not true by necessity. As he wrote in a letter of 1817 to the Bremen astronomer Heinrich Wilhelm Olbers (of Olbers' paradox fame): "Maybe in another life we shall attain insights into the essence of space which are now beyond our reach. Until then we should class geometry not with arithmetic, which stands purely a priori, but, say, with mechanics" [4, p. 177].

Gauss realized that if the question could be settled at all, it would require astronomical measurements over very large, stellar distances. According to a popular and oft-repeated story, his high-precision geodesic measurements of a triangle spanning three mountains in Germany were undertaken with the aim of testing the assumption of Euclidean geometry. The sides of the Brocken-Hohenhagen-Inselsberg triangle were approximately 69, 85, and 107 km. However, historians have shown that the story is nothing but a persistent myth [5].

Although Gauss seems to have been aware that information about the curvature of space could in principle be obtained from data of stellar parallaxes, he did not pursue this line of reasoning. In a letter to the German-Danish astronomer Heinrich C. Schumacher of 12 July 1831 [6, p. 270] he communicated the formula for the circumference of a circle of radius $r$ in the new geometry, stating it to be

$$\ell = \frac{\pi}{2} k \left[ exp\left(\frac{r}{k}\right) - exp\left(-\frac{r}{k}\right) \right]$$

The quantity $k$ is a constant to be determined observationally and which, in the Euclidean case, is infinite. Should space not be Euclidean, Gauss said, from an observational point of view we can only say that the curvature measure $k$ must be incredibly large. In response to a letter from Gauss, the German-Baltic astronomer Friedrich Wilhelm Bessel admitted that "our geometry is incomplete and should be supplied with a hypothetical correction that disappears in the case that the sum of angles in a plane triangle = 180°" [7, p. 493].

The true founders of non-Euclidean geometry were the Hungarian mathematician János Bolyai and the Russian Nikolai Ivanovich Lobachevsky, both of whom (contrary to Gauss) published their independent discoveries that a geometry different from and as valid as Euclid's is possible. Whereas Bolyai did not refer to the skies, the astronomy-trained Lobachevsky did. Although primarily a mathematician, as a young man he had studied astronomy and in the 1820s he served as director of the Kasan University Observatory.

Already in his 1829 paper in the *Kasan Messenger* Lobachevsky suggested that one consequence of his "imaginary" (or hyperbolic) geometry might be tested by astronomical means, namely, that the angle sum of a triangle is always less than 180° and the more so the bigger the triangle becomes [8; 9]. From astronomical data he concluded that the angle sum of the triangle spanning the Sun, the Earth, and Sirius deviated from the Euclidean value of 180° by at most 0".000372, evidently much less than the observational error. In fact, the true deviation was even less [10]. Nonetheless, Lobachevsky realized that while it could in principle be proved that astronomical space is non-Euclidean, it could never be proved to be Euclidean, and for this reason he tended to see his comparison as inconclusive.

In a later work, titled *Pangeometry* and translated into French in 1856, he argued that, assuming space to be hyperbolic, there must be a minimum parallax for all stars irrespective of their distances from the earth. This is contrary to Euclidean space, where the parallax tends toward zero as the distance increases toward infinity.

The ideas of non-Euclidean geometry pioneered by Gauss, Lobachevsky, and Bolyai circulated but slowly in the mathematical community. Only about 1870, after the theory had been presented in a more elaborate form by Eugenio Beltrami in Italy and Felix Klein in Germany, and also been disseminated by Hermann von Helmholtz, did they truly enter the world of mathematics. The new ideas then resulted in a revolution in geometry. Of particular importance was the work of the Göttingen mathematician and physicist Bernhard Riemann, who in a famous address of 1854 put the concept of curvature as an intrinsic property of space on a firmer basis [11; 12]. Importantly, his address led to the now standard distinction between three geometries of constant curvature. The three possibilities correspond to flat or Euclidean space (curvature constant $k = 0$), spherical space ($k = +1$), and hyperbolic space ($k = -1$).

According to Riemann, although there are any number of possible geometries, only these three have properties that make them candidates for the real space we live in: they are homogeneous and isotropic, and also invariant under rotation and translation. Riemann was the first to point out that, in the case of constant positive curvature, the traditional identification of a finite three-dimensional space with a bounded space is unwarranted. As he wrote, "if we assume independence of bodies from position, and therefore ascribe to space constant curvature, it must necessarily be finite provided this curvature has ever so small a positive value" [13, p. 36].

In his 1854 address Riemann only referred to astronomy in passing, pointing out that on the assumption of a constant-curvature space it follows "from astronomical measurements that it [the curvature] cannot be different from zero." Although he accepted the idea of just one physical space, he did not accept that the geometry of this space could be known a priori or with absolute certainty. Our experience about the physical world, he said, could be consistent with geometries of different kinds. Moreover, he left open the possibility that on a microphysical scale the curvature of space might vary, if in such a way that the averaged curvature over measurable distances becomes unappreciably close to zero. Unfortunately he did not elaborate.

## 2.2. From Zöllner to Schwarzschild

Although Riemann's emphasis on the possibility of an unbounded yet finite space implicitly addressed an old cosmological conundrum, it failed to attract interest among astronomers. It took nearly two decades before a scientist made astronomical use of Riemann's insight and then without making an impact on the astronomical community.

Karl Johann Friedrich Zöllner is today recognized for his contributions to astrophysics and, in particular, his pioneering work in astrophotometry [14; 15]. A skilled experimentalist and designer of instruments, in 1858 he invented an astrophotometer to measure the feeble light from stars and planets. In 1862 he moved to Leipzig, where he was appointed professor and established an astrophysical research program, the first of its kind. He may have been the first to use the name "astrophysics," which he introduced in 1865. In addition to his experimental work, he also made important studies of theoretical problems in astronomy and physics. These included electrodynamics, solar theory, sunspots, geomagnetism, and the theory of comets.

In his controversial book *Natur der Cometen* from 1872 Zöllner developed an electrical theory of comets and their tails that for a period was widely admired [19]. The book is not well known and has never been translated into English. In spite of its obscurity it, or rather one of the chapters in it, has an important position in the history of cosmology.

According to Zöllner [16], for two elementary particles of mass $m$ and charges $\pm e$ the attractive electrical force would exceed the repulsive force by a factor $(1 + \gamma)$. For the tiny quantity $\gamma$ he obtained the expression

$$\frac{1}{2\gamma} = \frac{e^2}{Gm^2} \cong 3 \times 10^{40},$$

from which $\gamma = 1.7 \times 10^{-40}$. Thus, the electric interaction between two neutral bodies would not be zero. A very small residual force would remain between them, and this residual electric force he identified with the gravitational attraction. This may have been the first recognition of the later so famous ratio between the gravitational and the electromagnetic interaction. The dimensionless and still unexplained number of the order $10^{-40}$ is today often referred to as either Weyl's or Eddington's (or sometimes Dirac's) number [17]. It only played a role in fundamental physics in the early part of the twentieth century, but can be traced back to Zöllner in about 1880.

*Natur der Cometen* included a chapter on "The Finitude of Matter in Infinite Space" in which Zöllner offered an original solution to Olbers' paradox in terms of a universe of constant positive curvature [18]. In his systematic discussion of the finite versus the infinite in the universe, he assumed, for the sake of discussion, that there is only a finite amount of matter in the world. He then argued that in an unbounded (and therefore infinite) Euclidean space any finite amount of matter would evaporate and dissolve to zero density in an eternity of time. Given the actual existence of matter of non-zero density, he concluded that either is space finite or the universe has only existed in a limited period of time. Unwilling to accept the latter hypothesis, he suggested that Riemann's geometry might provide the key that would unravel the secrets of the universe and dissolve the problems of a materially finite universe.

"It seems to me," he wrote [19, p. 308], "that any contradictions will disappear … if we ascribe to the constant curvature of space not the value zero but a positive value of the spatial curvature measure involves us in no way in contradictions with the phenomena of the experienced world if only its value is taken to be sufficiently small." In this way he made Olbers' paradox disappear without having to assume a limitation of either cosmic time or space. He also suggested that in a Riemannian universe all processes would occur cyclically, indeed that the universe itself would be cyclic.

Zöllner's innovative cosmological speculations attracted some attention in German philosophical circles, but were ignored by almost all physicists and astronomers. Generally speaking, during the nineteenth century problems of cosmology were of limited interest to astronomy, a science which primarily dealt with the solar system, the stars making up the Milky Way, and the enigmatic nebulae. In the spirit of positivism, the large majority of astronomers tended to conceive the universe at large as a field for philosophical study rather than scientific exploration. As long as the riddle of the nebulae – that is, whether or not the nebulae belonged to the Milky Way – remained unsolved, cosmology was bound to remain speculative, hence unscientific. From the perspective of this attitude, which was commonly held by astronomers in the Victorian era, there was little to recommend Zöllner's theory of a spatially closed universe.

Only a handful of astronomers in the late nineteenth century expressed any concern about the possibility of space being non-Euclidean, and none of them received inspiration from or even knew about Zöllner's pioneering work. The distinguished American astronomer Simon Newcomb contributed to the mathematical aspects of curved space by distinguishing between elliptic space and spherical space. He also had an interest in the potential application to astronomy, which he commented on from time to time. However, he did not seriously believe in astronomical space being curved.

Like Lobachevsky many years earlier, Newcomb pointed out that the hypothesis was testable, if this might be more in principle than in practice. As he wrote: "Unfortunately, we cannot triangulate from star to star; our limits are the two extremes of the earth's orbit. All we can say is that, within those narrow limits, the measures of stellar parallax give no indication that the sum of the angles of a triangle in stellar space differs from two right angles" [20, p. 7].

Newcomb's friend and correspondent, the philosopher, scientist, and polymath Charles Sanders Peirce, disagreed. Convinced that space must be non-Euclidean and most likely hyperbolic, he offered a series of detailed arguments of both a philosophical and astronomical kind [21]. Peirce's attempt to conceive celestial space as non-Euclidean was the most elaborate and serious one of the few such attempts in the late nineteenth century. Not only did he consider the effect of curved space on measurements of stellar parallaxes, he also did the same with respect to the proper motions of stars, stellar evolution, and the Doppler shifts in stellar spectra.

However, Peirce's ideas made no impact at all. He mainly communicated them in the form of letters or papers in the philosophical literature rather than publishing his arguments in journals read by astronomers and mathematicians. Although they were known to some American scientists, they failed to convince them. As Newcomb tersely wrote him in March 1892, "the task of getting the scientific world to accept any proof that space is not homoloidal [flat], is hopeless, and you could have no other satisfaction than that of doing a work for posterity" [22, p. 424].

In a lecture given on 9 August 1900 to the Astronomical Society in Heidelberg, 27-year-old Karl Schwarzschild discussed systematically how to determine the geometry of space from observations [23; 24; 25]. While in Euclidean space the parallax $p$ of a star infinitely far away is zero, in hyperbolic space there will be a minimal non-zero parallax that decreases with the curvature radius $R$ given by

$$R^2 = \frac{k}{K}, \quad k = \pm 1$$

In a triangle spanned by a star and the two positions of the Earth half a year apart in its orbit around the Sun we have

$$p = \pi - (\beta + \gamma) = \alpha - K\delta,$$

where the angle at the star is denoted $\alpha$ and the two angles at the positions of the Earth $\beta$ and $\gamma$. Thus, in the hyperbolic case the parallax of distant stars ($\alpha \cong 0$) will remain positive. If stars are observed with a zero parallax, meaning a parallax of the same magnitude as the error of observation, the error will give an upper limit to the numerical curvature. In the case of a distant star in spherical space ($K > 0$), the sum of $\beta$ and $\gamma$ will be greater than $\pi$ and so the parallax should be negative. If no stars are observed with $p < 0$, the error of observation will again give an upper limit to $K$. Schwarzschild estimated $p_{min} \cong 0".005$, from which he concluded that $R > 4 \times 10^6$ AU.

By applying arguments based on the parallax method and supplementing them with a method based on star counts, Schwarzschild [23, p. 345] arrived at the following conclusion: "One may, without coming into contradiction with experience, conceive the world to be contained in a hyperbolic (pseudo-spherical) space with a radius of curvature greater than 4,000,000 earth radii, or in a finite elliptic space with a radius of curvature greater than 100,000,000 earth radii, where, in the last case, one assumes an absorption of light circumnavigating the world corresponding to 40 magnitudes."

He saw no way to go further than this rather indefinite conclusion and decide observationally whether space really has a negative or positive curvature, or whether it really is finite or infinite. Nonetheless, from a philosophical point of view he preferred a closed universe. His reason was that "then a time will come when space will have been investigated like the surface of the earth, where macroscopic investigations are complete and only the microscopic ones need continue." Interestingly, some later cosmologists, including Eddington and Lemaître, expressed their preference for the closed universe models of general relativity in similar philosophical language.

Schwarzschild's paper of 1900 was not much noticed and he did not himself think of it as important. Eight years later his considerations were extended by another German astronomer, Paul Harzer, who developed a more detailed model of a closed stellar universe [26; 3]. Published in a mathematical journal, Harzer's model attracted little attention and failed to change the attitude of the astronomers. The main reason why the idea of curved space failed to catch the interest of astronomers was simply that they had no need for it.

### 2.3. Relativity Theory and Curved Space

Non-Euclidean geometry played no role in Einstein's restricted or special theory of relativity or in his first attempts to extend it into a theory of gravitation. Only in about 1913 did he realize the relevance of Riemann's ideas of differential geometry, which proved crucial in the construction of the covariant general theory in the fall of 1915. As well known, his theory predicted a curvature of space caused by massive bodies, which was verified in 1919 with the detection of the bending of starlight in the famous Eddington-Dyson solar eclipse expedition. Of course, this was a local curvature of space caused by the sun's gravitational field and not a proof that global space is positively curved.

The generally accepted picture of the universe about 1915, and one that Einstein roughly subscribed to, was a huge stellar system of an ellipsoidal form, with the density of stars diminishing with increasing distance from the center [27]. The dimensions of the system were thought to be of the orders 50,000 light years in the galactic plane and 5,000 light years towards the galactic poles. This picture of the Milky Way universe built upon the statistical investigations of authorities such as Jacobus Kapteyn in the Netherlands and Schwarzschild and his former teacher Hugo von Seeliger in Germany. It was accepted by a majority of astronomers, if not by all of them.

The material universe was usually considered a finite stellar system in the infinite Euclidean space and often identified with the Milky Way. What might be beyond the stellar system was left to speculation. It might be empty space or some ethereal medium, but in any case it was regarded as irrelevant from an astronomical point of view. Assuming that starlight was not absorbed by interstellar matter, the mass density of the Milky Way universe was estimated to be of the order $10^{-23}$ g/cm$^3$. The suggestion of Schwarzschild and Harzer of a closed space filled with stars had the advantage that it did away with the troublesome infinite empty space in which the stellar system was presumably embedded, but the idea made almost no impact on mainstream astronomy.

Einstein did not originally think of using his theory of general relativity in a cosmological context. The idea seems to have come from Schwarzschild, who in February 1916 informed Einstein that his relativistic field equations have a solution corresponding to a closed universe with an elliptic geometry of the type discussed by Newcomb [25]. Later the same year Willem de Sitter discussed the possibility of a relativistic description of the universe with Einstein. Only in the fall of 1916 did he seriously investigate the problem, which turned out to be more difficult than he had expected. He had to think deeply about the classical problem of boundary conditions until he realized that they were not needed.

The result of Einstein's thinking were the cosmological field equations of early 1917, including the cosmological constant $\Lambda$ that he thought was needed to keep the closed universe in a stationary state. His model of the universe was four-dimensional in space-time, and, satisfying the requirements of homogeneity and isotropy, with its metric being separable in the three space coordinates and the one time coordinate [28]. Although the curvature of space would vary locally in time and space in accordance with the distribution of matter, he considered spherical space to be a good approximation on a cosmological scale.

Einstein's theory resulted in definite formulae relating the mass and volume of the universe to the radius of curvature $R$, such as

$$M = 2\pi^2 \rho R^3$$

However, given the uncertainty of the average density of matter $\rho$ this was of little help. In correspondence with his friend Michel Besso of March 1917, he suggested that $R \approx 10^7$ light years, a value which was based on the estimate $\rho \approx 10^{-22}$ g/cm$^3$. Although this estimate was not far from the density that Kapteyn had suggested for the Milky Way, it was orders of magnitude greater than the value later obtained by Hubble (namely, $\rho \cong 10^{-31}$ g/cm$^3$). Some months later he repeated the suggestion in a letter to de Sitter, but he wisely decided not to publish it [29].

In an address on geometry and experience that Einstein gave to the Prussian Academy of Sciences in 1921 he distinguished between what he called "practical geometry" and "purely axiomatic geometry." He argued that while the first version was a natural science, the second was not. "The question whether the universe is spatially finite or not seems to me an entirely meaningful question in the sense of practical geometry," he said. "I do not even consider it impossible that the question will be answered before long by astronomy." Indeed, without this view of geometry, he continued, "I should have been unable to formulate the theory of [general] relativity" [30, p. 235-239]. According to Einstein, geometry did not in itself correspond to anything experienced. It would only do so if combined with the laws of mechanics and optics, or with other laws of physics.

Incidentally, the suggestion that Einstein made in 1921, that astronomical observations would soon reveal whether cosmic space is curved or not, turned out to be unfounded. Still in 1931, after he had accepted the expansion of the universe, he stuck to a closed universe, such as shown by his model of a cyclic universe from that year. But the following year he changed his mind. In the important model he proposed jointly with de Sitter, and to which I shall return in Section 3.4, he admitted that "There is no direct observational evidence for the curvature, … [and] from the direct data of observation we can derive neither the sign nor the value of the curvature" [31]. For reasons of simplicity, the Einstein-de Sitter model therefore described the universe without introducing a curvature at all. As far as the curvature of cosmic space was concerned, the Einsteinian revolution in cosmology did not change much.

# 3. RELATIVISTIC MODELS OF THE UNIVERSE

## 3.1. Observations

During the period from about 1910 to 1930 there was little connection between astronomy and fundamental physics as far as the structure of the universe was concerned. After Einstein introduced his relativistic theory of cosmology in 1917, the theory attracted attention among a small group of physicists, mathematicians, and astronomers, but for a while astronomical observations played only an insignificant role. As far as the astronomers were concerned, most of them disregarded the new and mathematically abstruse theory, continuing to chart the universe by means of observations and to relate their data to classical models of the stellar universe. It is a mistake to believe that cosmology anno 1925 was generally affected by the new theory of general relativity.

If there were a burning question in observational cosmology in the early part of the twentieth century, it concerned the location of the nebulae relative to the Milky Way system. It was an old question, going back to the famous philosopher Immanuel Kant in the mid-eighteenth century and later independently discussed by William Herschel in England [32; 2, pp. 75-83]. Some astronomers advocated a modernized version of Kant's view, namely, that the nebulae, and especially the spiral nebulae, were structures similar in size and shape to the Milky Way. This view was known as the "island universe" theory.

According to the alternative view, the Milky Way system was essentially the entire material universe, whereas the nebulae were relatively small structures located within its confines. At the turn of the century the latter view was favored by a majority of astronomers. "No competent thinker," said the British astronomer Agnes Clerke [33, p. 368], "can now, it is safe to say, maintain any single nebula to be a star system of coordinate rank with the Milky Way." However, the whole question of island universe versus the Milky Way universe remained unresolved, primarily because the distances to the far-away nebulae were unknown.

In 1918 the young Mount Wilson astronomer Harlow Shapley created a minor sensation when he proposed a monster Milky Way much larger than previously argued. In a letter to George Hale he described his stellar, "galactocentric" universe as an "enormous, all-comprehending galactic system … the diameter [of which] is some 300,000 light years in the plane" [34, p. 62]. For the thickness of the system, he estimated a value of 30,000 light years. Shapley found his immense galactic system to be incompatible with the island universe theory, for if the spiral nebulae were external galaxies comparable in size to the Milky Way they would have to be at inconceivably great distances.

On 20 April 1920 the two opposing views of the universe were discussed at the so-called "Great "Debate" meeting organized by the National Academy of Science in Washington D.C. The questions under discussion were the size of the Milky Way and the distribution of the spiral nebulae relative to it. Shapley's opponent Heber Curtis defended the picture of an island universe, which he described as follows: "The spirals are a class apart, and not intra-galactic objects. As island universes, of the same order of size as our galaxies, they are distant from us 500,000 to 10,000,000 or more, light years" [35, p. 303; 27]. As to the diameter of the Milky Way he favored the traditional maximum value of about 30,000 light years or one-tenth of Shapley's value.

The Great Debate did not lead to any consensus, but only increased the confusion until Edwin Hubble famously resolved the question by detecting Cepheid variables in the Andromeda Nebula. This allowed him to estimate its distance by means of the period-luminosity relation that had been known since about 1910. In Hubble's paper of 1925 he reported the distance to be about 930,000 light years, placing Andromeda well beyond the limits of even Shapley's Milky Way [36, pp. 713-715; 27]. He underestimated the distance by a factor of more than two, but that was only recognized in the early 1950s, when it led to a drastic revision of the length and time scales of the universe.

Hubble's discovery dramatically changed the attitude in the astronomical community in favor of the island universe theory. By the late 1920s Hubble's modernized version of Kant's old view had become accepted by most of the leading astronomers. Even before the discovery, this theory of the universe was gaining support from quite a different kind of observations, namely, the redshifts of the nebulae discovered by Vesto Melvin Slipher at the Lowell Observatory [37; 27].

Slipher first found that the spectral lines of the Andromeda Nebula were shifted towards the blue (a result of its local motion), but he soon realized that in general the nebulae exhibited redshifts. By 1917 he had determined redshifts for 21 spirals and, interpreting them as Doppler shifts, found their radial velocities away from the Sun to be between 300 km/s and 1,100 km/s. The discovery did not arouse much immediate response, but as more redshifts were found it became important to find the mechanism that caused the apparent recession of the spirals. It was in this context that the redshifts entered the debate of the structure of the universe, namely, as support of the island universe view. As early as March 1914, the Danish astronomer Ejnar Hertzsprung wrote to Slipher: "It seems to me, that with this discovery the great question, if the spirals belong to the system of the Milky Way or not, is answered with great certainty to the end that they do not" [34, p. 22]. However, it was Hubble's discovery and not Slipher's that settled the question.

Although few observational astronomers cared about or were even closely acquainted with Einstein's view of the closed universe, by the mid-1920s the situation began to change. Not only did a few mathematically inclined astronomers such as Eddington and de Sitter investigate the relativistic universe, Einstein's theory also began to appear relevant with regard to astronomical measurements. Slipher's redshifts were not alone in attracting attention to the tensor equations of general relativity.

In an important paper of 1926 on the classification of nebulae, Hubble obtained an average mass density of the universe of $1.5 \times 10^{-31}$ g/cm$^3$, which was much less than previous estimates. Rather than just reporting his result, at the end of the paper he inserted the density value in the expressions that Einstein had given for the curvature radius and mass of his closed universe. The results were $R = 2.7 \times 10^{10}$ parsecs and $M = 9 \times 10^{23}$ solar masses. Hubble thought that with larger and more powerful telescopes than even the Mount Wilson 100-inch reflector, "it may become possible to observe an appreciate fraction of the Einstein universe" [36, p. 724]. Important as it is that Hubble related his observations to Einstein's theory, it is also significant that he had his knowledge of the theory only from second hand. Rather than reading Einstein's paper, he had his information from a general textbook in physics written by the Austrian Arthur Haas.

### 3.2. Static Universe Models

As mentioned, in early 1917 Einstein published a seminal paper in the proceedings of the Prussian Academy of Sciences in which he applied his new theory of gravitation to a spatially closed and therefore finite universe [28]. His model universe was filled with a finite amount of homogeneously distributed matter, and in its time-dimension it was infinite, that is, static. "The curvature of space is variable in time and space, according to the distribution of matter, but we may roughly approximate to it by means of a spherical space," he wrote. Einstein found such a picture to be "logically consistent" and the one "nearest at hand" from the standpoint of general relativity. At the very end of his paper he added, significantly: "Whether, from the standpoint of present astronomical knowledge, it is tenable, will not here be discussed."

The astronomical knowledge that Einstein referred to indicated that the universe as a whole did not vary in time. For this reason he modified the field equations of general relativity by adding a term proportional to the metric tensor. The new term $\Lambda g_{\mu\nu}$ (or $\lambda g_{\mu\nu}$ in the original nomenclature) included a constant of proportionality which he originally referred to as just an "unknown universal constant." The new constant, $\lambda$ or $\Lambda$, would soon be known as the cosmological constant. With the quantity $\kappa$ denoting the Einstein gravitational constant ($\kappa = 8\pi G/c^2$), Einstein wrote the equations in a form close to the modern one:

$$R_{\mu\nu} - \frac{1}{2}g_{\mu\nu}R - \Lambda g_{\mu\nu} = -\kappa T_{\mu\nu}$$

As Einstein saw it, the constant was a property of space-time (rather than matter-energy) and necessary in order to keep the material universe static in spite of the attractive force of gravity. For this reason he placed it on the left side of the equation. Admitting that it was of an ad hoc character, he nonetheless found it necessary, for other reasons because he could then give an expression for the average density of matter in the universe. Einstein stated the following relations as characteristic for his closed and static model:

$$\Lambda = \frac{1}{2}\kappa\rho = R^{-2}, \quad V = 2\pi^2 R^3, \quad M = 2\pi^2 \rho R^3$$

As to the value of the new cosmological constant, in the Einstein model it had to be positive and exceedingly small in order not to spoil the agreement between general relativity (with $\Lambda = 0$) and planetary motions. While Einstein did not offer an estimate of its size, Willem de Sitter did: "Observations will never be able to prove that $\lambda$ vanishes, only that $\lambda$ is smaller than a given value. Today I would say that $\lambda$ is *certainly* smaller than $10^{-45}$ cm$^{-2}$ and is probably smaller than $10^{-50}$." He added, prophetically: "Maybe observations will one day provide a specific value for $\lambda$, but up to now I have no knowledge of anything pointing to this" [2, p. 133].

To Einstein's surprise, in a report to the Royal Astronomical Society of 1917 de Sitter proved that there exists another solution to the cosmological field equations, namely, one describing an empty universe with $\Lambda = 3/R^2$. Contrary to the modern understanding of the de Sitter solution, his model of 1917 was, like Einstein's, static and spatially closed. During the period 1917-1930 the two solutions or models were normally designated the A (Einstein) and the B (de

Sitter) solution. A small group of mathematically minded physicists and astronomers analyzed the properties of the two solutions, proposed modifications to them and compared them as candidates for the real universe. The primary aim of this work was to determine which of the two rival models of the static universe was the most satisfactory. Until the late 1920s the general attitude was that either the A or the B model was the correct one [38].

Whereas the conceptually simple A model was the one that attracted public attention, to physicists and astronomers the B model was no less interesting. Although de Sitter's model was devoid of matter, this was not necessarily seen as a problem, for the model could be regarded as a zero-density approximation to the physical universe of very low density. Moreover, the B model had some remarkable properties that made it attractive in connection with the nebular redshifts discovered by Slipher.

The redshift phenomenon was foreign to Einstein's model, but not to de Sitter's. In fact, as early as 1917 the Dutch astronomer pointed out that some of the predictions of his theory appeared to be related to Slipher's measurements of radial nebular velocities. The general redshift phenomenon, he wrote, "would certainly be an indication to adopt the hypothesis B in preference to A" [39, p. 28]. De Sitter's theory predicted a systematic displacement of the spectral lines toward the red, but he was careful to describe the corresponding radial velocities of the light sources as "spurious." It was, he explained, an effect of a particular space-time metric and not a result of a Doppler shift caused by the expansion of space. Indeed, still in the mid-1920s the concept of expanding space was nearly unthinkable.

There was in the period several attempts to relate the redshifts of the spirals found observationally to those predicted by de Sitter's theory [40; 34]. It was generally agreed that somehow the redshifts varied systematically with the distance, but there was neither observational nor theoretical agreement as to the form of the relation. Was it linear? Or was it perhaps quadratic, as in de Sitter's theory? In 1923 Weyl calculated that for relatively small distances the redshift $z = \Delta\lambda/\lambda$ would vary linearly with the distance $r$. The following year Ludwik Silberstein argued for a relation of the form $z \cong \pm r/R$, with $R = 6 \times 10^{12}$ AU. As shown by the double sign, his relation referred to blueshifts as well as redshifts.

However, astronomical data failed to support any definite relationship, whether linear or not. The main reason was that the distances to the spiral nebulae were little more but educated guesswork. Only in the aftermath of Hubble's discovery of 1925 did the situation improve, eventually leading to the linear redshift-distance law four years later. By that time cosmology was in a state of flux and the important problem was no longer to decide between two static world models, Einstein's A model and de Sitter's B model. A non-static model that somehow integrated the virtues of the two static models began looking more promising, perhaps even necessary.

### 3.3. The Expanding Universe

The story of the discovery of the expanding universe is often misrepresented as starting with Hubble's discovery in 1929 of a linear redshift-distance relation. However, in this case theory came before observation. Not only did Hubble not claim to have discovered the expanding universe, it was a discovery that relied crucially on theory and could not possibly have been done purely observationally. The possibility that the universe is in a state of expansion had been predicted several years before Hubble, first by Alexander Friedmann and then independently and more committedly by Georges Lemaître.

A professor of physics in St. Petersburg, Friedmann wrote in 1922 a paper that later would be regarded a classic in the cosmology literature but at the time attracted almost no attention [41; 36, pp. 838-843]. Analyzing systematically the solutions of the Einstein equations, he realized that there was a whole class of non-static, dynamical solutions corresponding to the radius of curvature varying in time. Assuming for simplicity homogeneity and isotropy – that is, the cosmological principle – he arrived at a set of simple differential equations that described the possible variations of the space curvature $R(t)$. For closed models, where $R(t)$ is a measure of the size of the universe, he wrote the equations (with $R' \equiv dR/dt$ and $R'' \equiv d^2R/dt^2$) as

$$\left(\frac{R'}{R}\right)^2 + 2\frac{R''}{R} + \frac{c^2}{R^2} - \Lambda c^2 = 0$$

$$3\left(\frac{R'}{R}\right)^2 + 3\frac{c^2}{R^2} - \Lambda c^2 = \kappa\rho c^4$$

Friedmann not only discovered a class of expanding world models, he also realized that some of the models included $R = 0$ at $t = 0$ in the past, or what he called a "creation of the world." Moreover, he investigated cyclical models where $R(t)$ increases from $R = 0$ to a maximum value and then decreases to $R = 0$. He seems to have had an emotional preference for models of the oscillating type.

Whereas Friedmann thus demonstrated the mathematical possibility of an expanding universe, he did not argue that the real universe belongs to this type. His brilliant investigation was primarily a mathematical exercise unconcerned with observations and physics. For example, although he was aware of the galactic redshifts, he did not find it relevant to mention them. George Gamow, who in his youth had studied under Friedmann, later wrote that his original theory "started with a 'singular state' at which the density and temperature of matter were practically infinite" [42, p. 141]. However, Friedmann did not deal with temperature, density or radiation at all. It is no less misleading to portray him as "the man who made the universe expand," as the subtitle of a biography reads [43].

In any case, for nearly a decade Friedmann's paper remained either unknown or unappreciated. Einstein may have been the only one who actually responded to his theory, which he considered to be a mathematical speculation of no relevance to the real universe. As late as 1929 he maintained that the universe was finite in space and with an infinite and constant time-coordinate. He was at that time also familiar with Lemaître's work on the expanding universe, but neither did this theory shake his confidence in a closed and static universe.

Unaware of Friedmann's work, in 1927 the Belgian physicist, astronomer, and priest Georges Lemaître published a paper in which he duplicated much of the mathematics of Friedmann's investigation of the closed universe [36, pp. 844-848; 40]. His equations for $R(t)$ were the same, except that he added a pressure term. With this term he stated cosmological energy conservation in the form

$$dE + pdV = d(pV) + pdV = 0,$$

where $V(t) = \pi^2 R^3$. Contrary to Friedmann, he argued from astronomical data for a particular model, namely, a closed universe expanding from a static Einstein state. He estimated the radius of this state to be about 270 Mpc. The data Lemaître referred to were primarily Slipher's nebular redshifts, which he interpreted as an effect of the expansion of space and not as a Doppler effect due to the travel of the nebulae through space: "The receding velocities of extra-galactic nebulae are a cosmical effect of the expansion of the universe." Let light be emitted by a nebula when the radius of the universe is $R_1$ and received when it has increased to $R_2$. As Lemaître demonstrated, the result would be a spectral shift given by

$$z = \frac{\Delta \lambda}{\lambda} = \frac{R_2 - R_1}{R_1}$$

He even derived an approximately linear relation of the form $v = kr$ between recessional velocities and distances, estimating the recession constant to $k \cong 625$ km/s/Mpc. Unfortunately he published the important paper in a somewhat obscure Belgian journal, with the result that until 1930 it remained almost completely unknown. It was only then, in the wake of Hubble's redshift-distance measurements, that it became recognized as a landmark paper in the history of cosmology. When it appeared in an English translation in 1931, important parts of the original paper were mysteriously left out. As has only recently become known, Lemaître was himself responsible for the slightly but significantly abridged translation [44].

As Lemaître was unaware of Friedmann, so Hubble was unaware of both Friedmann and Lemaître. His famous "Hubble law" of 1929 was the result of an observational research program aiming to find the correct relation between the redshifts and distances of spiral nebulae. With more and better data than previous workers in the field, Hubble showed that up to a distance of two megaparsecs (corresponding to a recessional velocity $v \cong 1,000$ km/s) the redshifts or Doppler-velocities varied roughly linearly with the distances. In other words, he established as an empirical law that $v = Hr$, with $H$ soon to be known as the Hubble constant. For the value of the new constant he obtained $H \cong 500$ km/s/Mpc, of the same order as Lemaître's value but much too small according to later knowledge.

Although Hubble's paper was predominantly observational, he suggested that the explanation of the recession of the galaxies might be related to the "de Sitter effect" of general relativity. It is important to recognize that he did *not* interpret the redshifts or "apparent velocities" as Doppler shifts caused by the galaxies actually receding from the observer. Nor did he suggest that space is expanding, such as Lemaître had done. As he emphasized in a letter to de Sitter of 1931, he was content having demonstrated an empirical correlation. "The interpretation," he wrote, "should be left to you and the very few others who are competent to discuss the matter with authority" [34, p. 192]. When Hubble passed away in 1953, he was still not convinced that the universe is in fact expanding.

So, who discovered the expanding universe? The standard answer may be Hubble, but a much better choice is Lemaître, the less known Belgian cosmologist and Catholic priest [45]. In fact, the cautious empiricist Hubble never claimed to have discovered the expansion of the universe, but only to have found an empirical law connecting the redshifts and the distances of the spirals. The myth of Hubble as the discoverer of the expanding universe is of later date. At the time the Hubble relation was seen as interesting, but far from revolutionary, and his paper of 1929 received only few citations. No one considered it a proof that the universe is expanding.

The momentous change in attitude from a static to an expanding universe only occurred in the early 1930s and then primarily as a result of the late recognition of the work of Friedmann and Lemaître. Eddington strongly endorsed Lemaître's model, disseminating it to the wider astronomical community and at the same time improving it. As he emphasized, the static Einstein universe characterized by a special value of the cosmological constant, $\Lambda = \frac{1}{2}\kappa\rho$, was inherently unstable: it would start expanding if an ever so slight disturbance caused $\rho$ to drop below $2\Lambda/\kappa$.

In a paper of 1930 he presented his version of Lemaître's model or what came to be known as the Lemaître-Eddington model (or sometimes the Eddington-Lemaître model). The essence was this: "The radius of space was originally 1200 light-years … [and] its present rate of expansion is 1 per cent in about 20 million years" [46, p. 765]. De Sitter too saw Lemaître's expanding universe as a revelation. Within a few years the revolution was completed, with a majority of leading astronomers and physicists accepting the new picture of the cosmos.

However, there were loose ends and unsolved problems. For example, what was the mechanism that caused the instability of the Einstein universe? Why did it produce an expansion and not a contraction (after all, the field equations are time-symmetric)? And was the Einstein state really a true beginning, or did it itself come from some previous state? In addition, a substantial minority of astronomers questioned the consensus view of the expanding space and suggested alternatives that explained the redshifts on the basis of a static universe, so-called "tired light" hypotheses. This class of hypotheses continued to be defended for several decades, but without being taken seriously by mainstream astronomers.

Even if one accepted the expansion, it did not mean that *space* was expanding in accordance with general relativity. In the period between 1933 and 1948 the non-relativistic alternative suggested by Arthur Edward Milne attracted much attention especially among British scientists [47; 2, pp. 169-172]. Milne's model is a reminder that there is no one-to-one correspondence between the expanding universe and Einstein's relativity theory of gravitation.

### 3.4. A Universe of Finite Age

The Lemaître-Eddington model, probably the most favored during the 1930s, was spatially finite and with a positive cosmological constant. It was temporally infinite in so far that the universe expanded asymptotically from the static Einstein world, which presumably had existed in an eternity. There was a gradual (logarithmic) beginning of the expansion, but no proper origin of the universe. On the other hand, in 1922 Friedmann had referred to the possibility of a "creation" in a singular state $R = 0$ at $t = 0$, and as soon as the Friedmann equations became generally known, models of this kind were considered, if only formally and somewhat reluctantly.

The idea of a big bang universe is usually and with good reason ascribed Lemaître, but Einstein had the same idea slightly earlier. However, to him (as to Friedmann) the state $R = 0$ was of a mathematical rather than physical nature. In the early spring of 1931 he proposed what Friedmann had done nine years earlier, namely, a closed cyclic model with $\Lambda = 0$ starting and ending in $R = 0$ and governed by the equation

$$\left(\frac{dR}{dt}\right)^2 = c^2 \frac{R - R_{max}}{R}$$

The model formally depicted a universe of finite age with a "big bang" as well as a "big crunch," but Einstein remained silent about these features, perhaps considering them to be mathematical artefacts [48]. At about the same time Lemaître published his idea of a finite-age universe beginning in the explosion of an original mass or what he called a "primeval atom" [49; 50]. Contrary to Friedmann, Einstein, and de Sitter, he was committed to this kind of universe, which he emphasized was a physical and not merely a mathematical model. The difference in attitude between Lemaître and other scientists is important. One can reasonably call it the first physical big bang model of the universe.

Inspired by the appearance on Earth of radioactive elements with lifetimes of the order ten billion years, Lemaître suggested that originally all matter in the universe was condensed into a kind of gigantic and highly radioactive atomic nucleus. "We could conceive," he wrote in a now-famous note of 9 May 1931, "the beginning of the universe in the form of a unique atom, the atomic weight of which is the total mass of the universe … [and which] would divide in smaller and smaller atoms by a kind of super-radioactive process" [51].

Thus, at $t = 0$ the universe already existed in the shape of the primeval atom, the radius of which he estimated to be about one astronomical unit. The matter density corresponded to that of an atomic nucleus, of the order $10^{15}$ g/cm$^3$. At least in principle, such a hypothetical superatom was comprehensible and would, immediately after its disintegration, be subject to the laws of physics. On the other hand, Lemaître insisted that it was physically meaningless to speak of time before the initial explosion. Whereas he considered the primeval atom to be real, he denied that the cosmic singularity $R = 0$ formally turning up in the equations at $t = 0$ could be ascribed physical reality. As he expressed it, somehow nature would find a way of avoiding the "annihilation of space."

In Lemaître's scenario the universe expanded in three phases, which he characterized as follows: (i) A rapid, inflation-like expansion in which the primeval atom was broken down into "atomic stars." (ii) A period of slowing-down called the "stagnation phase." (iii) A phase of accelerated expansion corresponding to the present era. In the cosmic future the acceleration would continue indefinitely.

The cosmological constant played a crucial role in the theory, in particular because it controlled the length of the stagnation phase and hence the age of the universe. With $\Lambda_E$ denoting the Einstein value he wrote the constant as

$$\Lambda = \Lambda_E(1 + \varepsilon) = \frac{\rho\kappa}{2}(1 + \varepsilon),$$

Its physical meaning was that it provided a measure of the length of the stagnation phase. By adjusting the value of ε he could obtain an age of the universe far higher than the Hubble time and in this way avoid the age paradox (see Section 4).

Lemaître realized that his scenario of the exploding universe might appear artificial and unconvincing so long that he could not provide any evidence for the postulated initial explosion. He argued that there was in fact such evidence in the form of the cosmic rays, which he conceived as fossils from the cosmic past. According to his hypothesis the cosmic rays were not direct products of the original explosion but had their origin in the early formation of stars some ten billion years ago. As he phrased it, "cosmic rays would be glimpses of the primeval fireworks of the formation of a star from an atom, coming to us after their long journey through free space" [50, p. 31]. However, in spite of much work on the subject, some of it done in collaboration with the Mexican physicist Manuel Sandoval Vallarta, he failed to convince the physicists specializing in cosmic rays. Indeed, as the knowledge of the rays improved, it became increasingly clear that they could not be explained as the result of a singular explosive act in the past. Most of the cosmic rays turned out to have a more local origin, either solar or galactic.

Whereas Lemaître's theory of the expanding universe was received very favorably in the early 1930s, responses to the primeval atom hypothesis were quite different, typically characterized by dismissal or neglect. Eddington, for one, would have nothing to do with it. One critic called it "a brilliantly clever *jeu d'esprit*" and according to the Canadian astronomer John Plaskett it was "the wildest speculation of all … an example of speculation run mad without a shred of evidence to support it" [2, p. 155; 52]. At one occasion Hubble considered Lemaître's explosion model which, he showed, could be brought into agreement with observational data if a particular value of the cosmological constant were chosen. But even so the density would have to be suspiciously high (~ $10^{-26}$ g/cm$^3$) and the size of the universe suspiciously small (~ $4.7 \times 10^8$ light years). He consequently found it an unattractive model.

Generally speaking, big bang solutions of the Friedmann equations were rarely taken seriously in the 1930s and 1940s, and especially not if they were provided with a physical interpretation, as was the case with Lemaître's theory. Although a few physicists referred to the theory, no one developed it and by 1940 it seemed to have come to a dead end. It may be relevant at this point to refer to another and well known early model that formally was also of the big bang type, namely, the model proposed in 1932 jointly by Einstein and de Sitter [36, pp. 849-851]. The model was parsimonious in the sense that as many parameters as possible were assumed to be zero (pressure, space curvature, and the cosmological constant). It then follows from the Friedmann equations that the matter density is

$$\rho = \frac{3H^2}{8\pi G} = \frac{3}{8\pi GT^2},$$

where $T = 1/H$ is the Hubble time. With the accepted value of $H = 500$ km/s/Mpc it gives a density of a not unreasonable order, $\rho = 4 \times 10^{-28}$ g/cm$^3$. It further follows from the Einstein-de Sitter model that the scale factor $R(t)$ varies as

$$R(t) = \alpha t^{2/3} + \beta,$$

where $\beta$ is an arbitrary constant that can be taken to be zero. The brief paper of Einstein and de Sitter is of some importance because it was the first cosmological model assuming a flat and therefore infinite universe. It also implied an abrupt beginning in a singular state $R = \beta = 0$ only 1.2 billion years ago. Remarkably, the two distinguished authors passed over both features without even mentioning them. The scale factor varies with time as $R \sim t^{2/3}$, but the expression did not appear in the paper. In the 1930s the model was scarcely noticed and neither Einstein nor de Sitter took it very seriously. But later on it came to be seen as the prototype of big bang models.

Theories of the big bang type survived World War II, but only barely so. In the 1940s Lemaître's exploding universe was revived by George Gamow in the United States, whose version was developed independently and in some respects was substantially different. A pioneer nuclear physicist, Gamow's ambitious aim was to explain how the chemical elements were built up by nuclear reactions in the early universe in accordance with the roughly known cosmic distribution of the elements [53, pp. 81-141]. By following this approach of "nuclear archaeology" he hoped to get insight in the hot and dense inferno that he assumed was characteristic of the beginning of the expansion.

As early as 1940 Gamow described in a popular book "the creation of the universe from a primordial superdense gas" [54, p. 201], but only six years later did he develop his idea into a proper theory based on a beginning in a hot soup of neutrons of $\rho = 10^7$ g/cm$^3$ and $T = 10^{10}$ K. With $k$ designating the curvature parameter Gamow used the Friedmann equations in the form

$$\frac{1}{R}\frac{dR}{dt} = \sqrt{\frac{8\pi G\rho}{3} - \frac{kc^2}{R^2}}$$

From observational data he argued that the density term was smaller than the curvature term and consequently that $k = -1$. Contrary to Lemaître and most other cosmologists, he thus adopted an open and ever expanding universe. However, what really mattered to him were the nuclear reactions in the earliest phase of the expansion.

Another feature of Gamow's theory deserves mention. Rather than thinking of a truly primordial state that began expanding, he speculated that it was the result of the contraction of an earlier universe that had existed for ever [48]. What he called the "big squeeze" had given rise to the big bang. Gamow thus favored a bouncing, non-cyclical model. However, he realized that it was a speculation and that from an empirical point of view one might forget about the possibly state before $t = 0$.

Together with his student and collaborator Ralph Alpher, in 1948 Gamow developed his theory into the first version of what can reasonably be called a hot big bang theory. The basic mechanism was neutron capture combined with the decay of neutrons into protons and electrons. In this way Gamow and Alpher found that within half an hour all the nuclei of the elements could be formed in rough agreement with the abundance curve known empirically. The theory – sometimes taken to be the beginning of modern big bang cosmology – is known as the "αβγ theory," where the letters α and γ allude to Alpher and Gamow, respectively [36, pp. 864-865]. The third author of the paper was the famous German-American physicist Hans Bethe (β), who only entered nominally.

## 4. BIG BANG VERSUS STEADY STATE

### 4.1. Cosmology About 1950

For a few years following the αβγ paper of 1948, Gamow's research program in early-universe cosmology progressed, but then it came to an almost complete halt and was only resumed about a decade later. As Gamow and Alpher pointed out, at the high temperatures necessary for thermonuclear reactions the radiation density $\rho_r$ dominates over the matter density $\rho_m$. With expansion and cooling $\rho_r$ will decrease faster than $\rho_m$ and the universe will eventually become dominated by matter. In the first phase the densities will vary with time as

$$\rho_r \sim t^{-2} \text{ and } \rho_m \sim t^{-3/2},$$

and in the second,

$$\rho_r \sim t^{-4} \text{ and } \rho_m \sim t^{-3}$$

In the fall of 1948 Alpher and Robert Herman, another of Gamow's collaborators, calculated that the present radiation density was about $10^{-32}$ g/cm$^3$ as compared to $\rho_m \cong 10^{-30}$ g/cm$^3$. By means of the Stefan-Boltzmann law this "corresponds to a temperature now of the order 5 K." They further pointed out that "This mean temperature for the universe is to be interpreted as the background temperature which would result from the universal expansion alone" [55, p. 774]. Strangely from a later point of view, the prediction of a cosmic microwave background failed to attract the interest of physicists and astronomers. Ten years later it was effectively forgotten and only in 1965 did it reappear under very different circumstances.

The thermonuclear calculations of Gamow, Alpher, and Herman were at first promising, in so far that they resulted in a reasonable amount of helium in the universe, between 25% and 35% by weight. On the other hand, at the time the content of helium in the universe was only known very roughly and consequently the prediction was of limited value. More seriously, sustained efforts to find nuclear reaction mechanisms that allowed for the primordial formation of heavier elements failed miserably.

Latest by 1952 the "mass gap problem" was a reality, meaning that it proved impossible to produce nuclei of mass number $A > 5$ under the circumstances of the early universe. With the emergence in 1957 of the successful B$^2$HF theory (so named after its originators Margaret and Geoffrey Burbidge, Fred

Hoyle, and William Fowler), the attempt to explain element formation by primordial processes seemed much less appealing [56]. Since the $B^2HF$ theory gave a satisfactory explanation of almost all the elements (except helium and deuterium) based on nuclear reactions in stars and novae, there seemed to be no reason to involve processes in a hypothetical past of the universe. At least, this was the attitude of almost all astrophysicists.

By that time work on the Gamow-Alpher-Herman version of hot big bang theory had ceased. The theory was well known, but not much appreciated. Most astronomers and physicists found it to be unconvincing and lacking in testable predictions. They ignored or were unaware of the prediction of a cold microwave background, and, as mentioned, there were not as yet reliable observations of the cosmic helium abundance with which the helium calculations could be compared. The first glimpse of solid evidence in favor of Gamow's theory came in 1961, when Donald Osterbrock and John Rogerson estimated a helium content of 32%. Seen in retrospect, the Osterbrock-Rogerson paper was "the first well-documented proposal for a relation between the [big bang] theory and the observational evidence of a fossil from the early universe" [67, p. 59].

In the 1950s cosmology was not only a very small research area but also one that lacked disciplinary unity and a shared paradigm [92]. Even worse, it was still a matter of some debate whether physical cosmology could be counted as a proper science at all. As to size, according to *Physics Abstracts* the number per year of scientific publications in cosmology varied between 30 and 40 in the period around 1950. Ten years later the number had not increased. As to status, most astronomers were reluctant of admitting cosmology as a genuine branch of their science, primarily because of its unsettled theoretical foundation and the lack of connection to solid astronomical data. The attitude of most physicists was no more positive. In so far that cosmology was cultivated as a science (which it was, of course) the relativistic expanding universe was generally accepted, for example in the version of the Lemaître-Eddington model. But there were several alternatives, including Milne's theory, the new steady state theory, and various non-relativistic ideas of a static universe based upon tired-light and similar hypotheses.

The intellectual climate of ambiguity furnished a fertile soil for discussions of a methodological and foundational nature, which is characteristic of a science in its pre-paradigmatic phase. There were so many theories and models, and so few observations of the one and only universe, so how could one rationally prefer one model over another? It was common at the time to argue that cosmology involved "personal taste" and that the choice between models was unavoidably philosophical or aesthetic in nature. As the Swedish theoretical physicist Oskar Klein phrased it, it was all "a matter of taste" [53, p. 222]. According to the eminent astronomer Walter Baade, cosmology was "a waste of time" [57, p. 205].

All scientists interested in the field agreed that the basic weakness was the lack of relevant observations to distinguish between the bewildering number and variety of cosmological models. It was said, and only half in jest, that cosmology was a science that rested on only two and a half facts. The two facts were the darkness of the night sky (Olbers' paradox) and Hubble's empirical law, and the half fact was the expansion of the universe.

A contributing reason for doubting the cosmological models based on general relativity was that most of them predicted an impossibly low age of the universe [53, pp. 73-79]. For Friedmann models with $\Lambda = 0$ any non-zero density value will result in an age $t^*$ smaller than the Hubble time $T$:

$$t^* = \alpha T, \quad \alpha < 1$$

For example, $\alpha = 2/3$ according to the Einstein-de Sitter model. Only for an empty universe will $\alpha = 1$, and even that does not help. The problem was that the accepted Hubble time was close to 1.8 billion years, whereas the stars and even the Earth were known to be considerably older. It goes without saying that the universe cannot be younger than its constituent parts, hence the "age paradox." Until the mid-1950s there were various attempts to avoid the problem, but none of them were considered quite satisfactory. One possibility was to reintroduce $\Lambda > 0$, as in Lemaître's model, which was the main reason why the otherwise unpopular cosmological constant was kept alive in the period.

Another possibility would be to question the authority of Hubble's value $H \cong 500$ km/s/Mpc, but until 1952 no one thought it could be seriously wrong. The change in attitude only occurred when Baade announced from a recalibration of the period-luminosity relation that the Hubble time must be increased to at least 3.6 billion years. By the mid-1950s the accepted Hubble time had grown even bigger, to about 6 billion years. Although the age problem had eased, it had not yet disappeared. Thus the Earth was known to be 4.55 billion years old, that is, a little older than the age of the Einstein-de Sitter universe. Of course, the stars were even older, but many astronomers preferred to put the blame for the discrepancy on the admittedly uncertain stellar models rather than consider it a real problem for the evolutionary universe [58].

## 4.2. The Steady State Universe

The age paradox was one motive for the steady state alternative proposed in 1948 by Fred Hoyle, Hermann Bondi, and Thomas Gold [53]. It was published in two different versions, one by Hoyle and the other jointly by Bondi and Gold. Although the two versions differed considerably in methodology and style, they built on the same premises and led to the same conclusions. For this reason I shall treat them as just a single theory.

The three physicists agreed that the existing relativistic cosmology was unsatisfactory and needed to be replaced by a better theory of the universe. Apart from the age problem they also felt that the standard evolution theory was methodologically objectionable because it was not a single and testable theory, but rather a supermarket of different models. It was so wide that it could accommodate almost any observation, and hence had little real predictive power. In addition, the three physicists much disliked the notion of a beginning of the universe, which they thought was plainly unscientific. It was in this context that Hoyle coined the memorable term "big bang," which he first used in a BBC broadcast of 28 March 1949, contrasting the new steady state theory to theories based on "the hypothesis that all the matter in the universe was created in one big bang at a particular time in the remote past" [59]. The name did not initially catch on and was widely used only since the 1970s, after the theory of an explosive origin of the universe had been vindicated.

As indicated by its name, the essence of the new steady state theory was the assumption of an unchanging yet dynamic universe. Interestingly, this idea was considered as early as 1931 by none other than Einstein, shortly after he became convinced that the universe is in a state of expansion [60]. In an unpublished and undated manuscript probably dating from January 1931 he explored the possibility that the cosmological constant appearing in his field equations might be the source of a continuous formation of new matter. As he wrote, "By setting the λ–term, space itself is not empty of energy." However, he came to the conclusion that his steady state model of an expanding universe with constant matter density was untenable and decided to shelve the manuscript.

Contrary to other critics of the evolution theory based on general relativity, Hoyle, Bondi, and Gold fully accepted Hubble's recession law, the expansion of the universe, and the cosmological principle. But they extended the latter uniformity principle into what Bondi and Gold called the "perfect cosmological principle," namely, that the large-scale appearance of the universe is the same at any location *and* at any time. The new principle implied an eternal universe and thus eliminated the age paradox as well as the conceptual problems associated with a beginning of the universe. On the other hand, the matter density in an expanding universe decreases in time, apparently contradicting the perfect cosmological principle. According to Hoyle, Bondi, and Gold, it followed as a consequence of the stationarity principle combined with the expansion that matter must be continually created throughout the universe. The rate of creation would secure that the average density of matter remained constant.

The claim of spontaneous matter creation on a cosmic scale was the most radical and controversial part of the new steady state theory, but it was also unavoidable. The fact that it violated one of the most sacred laws of physics, the conservation of mass-energy, was not considered a mortal sin by advocates of the theory. After all, has energy conservation been proved experimentally to an indefinite accuracy? In his textbook *Cosmology* of 1952, Bondi wrote: "The principle resulting in greatest overall simplicity is then seen to be not the principle of conservation of matter but the perfect cosmological principle with its consequence of continual creation … [which] is the simplest and hence the most scientific extrapolation from the observations" [61, p. 144].

The nature of the new matter did not follow from the theory, but it was generally assumed to be hydrogen atoms or possibly neutrons. What did follow was the creation rate of matter, which in a space element $\Delta V$ was shown to be

$$\frac{dm}{\Delta V dt} = 3\rho H \cong 10^{-43}$$

in units $g/s/cm^3$. Corresponding to three new hydrogen atoms per cubic meter per million years, this was a rate far below direct experimental testing. However, indirectly the hypothesis was testable. Whereas the Hubble constant $H$, according to the relativistic theory, is slowly decreasing in cosmic time, in the steady state theory it is a true constant.

Given the semi-qualitative, almost philosophical foundation of the steady state theory, it is remarkable that it led to several unique and precise predictions, in this respect proving superior to the class of relativistic evolution theories. Apart from the creation rate the theory also led to a definite value of the constant density of matter in the universe, which happened to be exactly the same as the critical density in the Einstein-de Sitter model, $\rho = 3H^2/8\pi G$. As to the metric and the expansion of space, it follows from the constancy of $H = R'/R$ that the scale factor increases exponentially:

$$R(t) = R_0 \exp(Ht) = R_0 \exp(t/T)$$

Here the Hubble time $T$ is just a characteristic time scale that has nothing to do with the age of the universe (which is, of course, infinite). Moreover, since the rate of creation is constant and affected by the spatial curvature $k/R^2$, it follows that $k = 0$. The steady state space is thus flat and infinite, another feature it had in common with the Einstein-de Sitter space. From the mid-1950s it became common to characterize the expansion in terms of the dimensionless deceleration parameter $q_0$, which in the case of the steady state model is

$$q_0 \equiv -\left(\frac{d^2R/dt^2}{RH^2}\right)_0 = -1$$

Contrary to the relativistic theories, according to the steady state theory there are old and young galaxies in any large volume of space, galaxies being formed continually by accretion of new matter. The fraction of galaxies that are older than a certain age $t$ turns out to be $\exp(-3Ht)$ and the average age of galaxies to be $T/3$. Thus, steady state theory anno 1950 predicted an average age of galaxies of about 600 million years.

A few years after the steady state theory was announced, it gave rise to a heated controversy of both a scientific and a philosophical nature [53]. The controversy was most visible and noisy in England, whereas the new theory attracted much less attention elsewhere. American astronomers preferred to ignore it. The attitude of many mainstream cosmologists, including Lemaître and Gamow, was to dismiss it as artificial and speculative, scarcely worth a serious study. In England it attracted some support but also met with strong opposition because of the hypothesis of continual creation, which many scientists found to be preposterous.

Among the early British converts to the new cosmology was William McCrea, who developed Hoyle's version of the theory in such a way that it did not rely on matter creation as a primary postulate [62]. According to McCrea, the creation process was a consequence of space being endowed with a uniform negative pressure given by

$$p = -\rho c^2$$

However, because the pressure was completely uniform it would have no direct physical significance. The idea of a negative pressure or zero-point energy in empty space had earlier been proposed by Lemaître and it would later play an important role in cosmology (see Section 5).

The methodological issues that were discussed during the cosmological controversy involved not only physicists and astronomers, but also philosophers. The steady state theory was eminently falsifiable, while the relativistic evolution theory was not, and Bondi and his allies tended to see this feature as a strong argument for the new theory. They found support in the philosophy of Karl Popper, according to whom a highly falsifiable theory should be given precedence over less falsifiable alternatives. As Bondi said in a debate with an opponent of the steady state theory: "It is the purpose of a scientific hypothesis to stick out its neck, that is to be falsifiable. It is because the perfect cosmological principle is so extremely vulnerable that I regard it as a useful principle… I regard vulnerability to observation as the chief purpose of any theory" [63, p. 45].

Of course Bondi and other supporters of the steady state theory realized that falsifiability, although a methodological virtue, cannot be a criterion of truth. A highly falsifiable theory may well turn out to be wrong. It was generally agreed that the verdict concerning the truth of the cosmological models had to come from observations and not from philosophical arguments. And this is what happened.

### 4.3. Testing Cosmological Models

The challenge of the steady state theory was advantageous to the development of cosmology in the sense that it induced researchers to focus on the theory's sharp predictions and compare them with new observational results. The tests during the period ca. 1950-1965 were primarily aimed at distinguishing between the steady state model and the evolutionary models based on general relativity. Among the latter class, models of the big bang type played a role only at the end of the period. While some of the tests were of a theoretical or methodological nature – some of them in the form of thought experiments – others involved measurements and provided cosmology with a much needed basis of relevant observations. By 1965 cosmology was no longer a science based on merely two and a half facts. The most important of the empirical tests involved the following methods or phenomena:

- Nucleosynthesis of helium and heavier elements
- Redshift-magnitude relationship
- Angular diameter-redshift relationship
- Radio-astronomical source counts
- Cosmic microwave background

Of these, it was in particular the latter two that proved the steady state to be untenable and paved the way for the new hot big bang consensus model.

Whereas the successful stellar theory of nucleo-synthesis implicitly provided support for the steady state theory, it was unable to account for the amount of helium in the universe that in the early 1960s was, for the first time, reliably estimated to be about 30%. On the other hand, the figure agreed with the calculations based on Gamow's explosion theory. The idea behind the redshift-magnitude method had roots in the 1930s. Hubble's redshift-distance relation, expressed in terms of the apparent and absolute magnitudes of the galaxies ($m$ and $M$, respectively), can be written in a form that takes into account the evolution of the universe. The approximate expression is

$$m = M - 5\log(cz) + 1.086(1 - q_0)z,$$

where $z$ is the redshift. For data that extend to very large distances the relation provides information about the deceleration parameter $q_0$ and then about the geometry of space. The method was easy in principle, but very difficult in practice.

Allan Sandage at the Mount Palomar Observatory was convinced that the cosmological problem could be solved observationally and that it was basically about determining two numbers, $q_0$ and $H$ [64]. He was no less convinced that the steady state theory was all wrong. Analyzing data from 474 galaxies with redshifts up to $z = 0.2$, in 1956 he and his collaborators announced $q_0 = 2.5 \pm 1$, a result clearly inconsistent with the steady state theory. Although other measurements of this type gave lower results, none of them came close to $q_0 = -1$. Nonetheless, the values for $q_0$ were inaccurate and the method did not yield results that unambiguously ruled out the steady state theory, at least not in the eyes of the supporters of the theory. The evidence that the redshift-magnitude observations provided against the theory decreased its credibility, but it allowed it to survive.

More serious was the challenge from the new science of radio astronomy, which in the late 1950s was extended into the realm of cosmology [65]. Martin Ryle and his group in Cambridge counted the number $N$ of radio sources with a flux density larger than $S$. Assuming that the sources are uniformly distributed in a static flat space, the two quantities are related as

$$\log N(\geq S) = -1.5 \log S + \text{const.}$$

Cosmological models with different geometries and expansion rates will lead to different predictions, corresponding to number count results in a log$N$-log$S$ plot that can be compared with a straight line with slope $-1.5$. In particular, the steady state model predicts that all radio sources must lie beneath this line.

In 1955 Ryle concluded that the main part of the sources corresponded to a line of slope $-3$, strongly disagreeing with the steady state prediction. However, the conclusion was premature and contradicted by results obtained by Bernard Mills and his group of radio astronomers in Sydney. The Australian results indicated a slope of $-1.8$, which subsequently was raised to $-1.65$. For a while the method seemed inconclusive, but with more and better data a consensus value did appear.

In 1961 Ryle presented improved data that clearly disagreed with the steady state theory, and this time the result remained stable and was confirmed by the Sydney group. Two years later Ryle had narrowed down the slope to $-1.8 \pm 0.1$, in excellent agreement with the result $-1.85 \pm 0.1$ obtained by the Sydney group. Although the new consensus among radio astronomers did not kill the steady state theory, it left it seriously wounded. It indirectly gave strong support to relativistic evolution models, unfortunately without being able to single out the best among these models.

As is well known, the deathblow came with the celebrated discovery of the cosmic microwave background in late 1964, followed by a series of events that is thoroughly documented by historians and the actors themselves [66; 67; 68]. A brief summary follows. The discovery – or better, observation – of the background radiation was serendipitous, in the sense that when Arno Penzias and Robert Wilson made their radio-astronomical measurements they were not thinking of cosmology at all. All they found was a puzzling excess temperature in their antenna that somehow seemed to be of cosmic origin. Meanwhile Robert Dicke and his former student James Peebles were examining the consequences of a very hot and dense past of the universe, unaware that the question had been previously investigated by Gamow, Alpher and others. On Dicke's suggestion, Peebles calculated the properties of the relic radiation assumedly caused by what they, adopting a name suggested by John Wheeler, called the primeval "fireball." In early 1965 they were also unaware that Alpher and Herman had predicted a present microwave background of $T \cong 5$ K as early as 1948.

When Dicke and his collaborators in Princeton came to know of the excess temperature measured by Penzias and Wilson they quickly realized that it was due to the background radiation left over from the original fireball – they did not yet speak of a "big bang." The result was the famous publication in *Astrophysical Journal* of two companion papers, one by the Princeton group and the other by Penzias and Wilson. The cosmic microwave background had been discovered at $\lambda = 7.3$ cm and $T = 3.5 \pm 1.0$ K, and the following year it was confirmed by Roll and Wilkinson at λ = 3.2 cm and $T$ = 3.0 K.

The discovery agreed beautifully with the now revived big bang theory, but not at all with the steady state theory. It completed the killing of it, although not with the result that Hoyle and his allies immediately abandoned their favorite conception of the universe. On the other hand, revised versions of the steady state theory were marginalized and scarcely taken seriously in the new climate dominated by the hot big bang consensus theory. It remains to be said that Hoyle, in collaboration with Jayant Narlikar and Geoffrey Burbidge, continued to fight the big bang orthodoxy and in the 1990s suggested an alternative in the form of the so-called QSSC or quasi-steady-state cosmology [69; 70]. The theory was able to explain the microwave background and much more, but it was also considered baroque and hopelessly ad hoc. It made almost no impact at all on mainstream cosmology and is today largely abandoned.

# 5. DARK ENERGY BEFORE DARK ENERGY

## 5.1. The Vacuum, Classical and Quantum

In the late 1990s observations of supernovae made by two research teams (SCP and HZT) led to the surprising conclusion that the universe, although critically dense, is accelerating. The consensus interpretation of the data was that the acceleration is driven by a new and strange form of vacuum energy associated with the cosmological constant. The discovery of "dark energy" – a name dating from that time – is today seen as one of the most important discoveries of modern cosmology, matching in importance even the discovery of the cosmic microwave background (but not, perhaps, the discovery of the expanding universe). In 2011 the Nobel Prize in physics was awarded to Saul Perlmutter, Brian Schmidt, and Adam Riess, the researchers who first recognized the accelerating universe and the dark energy blowing it up. To trace the prehistory of this concept, one has to look at two conceptual sources that for about half a century lived separate lives. One is the cosmological constant and the other is the zero-point energy of the quantum vacuum.

Although empty space or vacuum was discovered experimentally in the seventeenth century by Torricelli, Boyle and others, at about 1900 few physicists believed that empty space was really empty. It was filled, they thought, with an ethereal, non-material medium, which was dynamically active and quite different from nothingness. According to the British physicist Oliver lodge, the ether was incompressible and a reservoir of an immense amount of potential energy. In 1907 he estimated the energy density of the ether to be about $10^{32}$ erg/cm$^3$ = $10^{25}$ J/cm$^3$, which by $E = mc^2$ is equivalent to $10^{11}$ g/cm$^3$ – "something like ten-thousand-million times that of platinum" [71]. No wonder that some modern physicists have seen in the modern form of vacuum energy a resurrection of the classical electromagnetic ether that was so popular in the Victorian era [72]. Several years before the discovery of the cosmic microwave background, Dicke [73, p. 29] wrote: "One suspects that, with empty space having so many properties, all that had been accomplished in destroying the ether was a semantic trick. The ether had been renamed the vacuum."

The concept of zero-point energy is known by all physicists, but it is less well known that it originated more than a decade before quantum mechanics. In 1911 Max Planck suggested that the average energy of a harmonic oscillator of frequency ν followed the expression

$$E_n = (n + \tfrac{1}{2})h\nu, \qquad n = 0,1,2,\ldots$$

rather than $E_n = nh\nu$. What soon became known as zero-point energy remained controversial until the mid-1920s when it turned up in molecular spectroscopy and was justified theoretically by the new quantum mechanics. However, whereas the zero-point energy of material objects (oscillators and rotators) was vindicated, most physicists denied the reality of zero-point radiation energy in free space.

In a paper of 1916, Walther Nernst defended the unorthodox idea that the zero-point energy was valid even for the radiation filling up space in the absence of matter [74]. Introducing a cut-off maximum frequency $\nu_m$ he found the total energy density of the ether to be

$$\rho = \int_0^{\nu_m} \frac{8\pi h}{c^3} \nu^3 d\nu = \frac{2\pi h}{c^3} \nu_m^4$$

Somewhat arbitrarily taking $\nu_m = 10^{20}$ Hz, Nernst obtained $\rho = 1.5 \times 10^{23}$ erg/cm$^3$ or about 150 g/cm$^3$, which made him describe the vacuum energy as "enormous." He also showed that if zero-point radiation enclosed in a container is compressed, neither its energy density nor its spectral distribution will be affected. To my knowledge, this is the first recognition of an invariant energy density, a property that much later would be seen as characteristic of dark energy.

Nernst's innovative idea attracted little attention, except that it was reconsidered by Wilhelm Lenz in an interesting paper of 1926 in which he investigated the thermodynamic properties of the static Einstein universe. However, Lenz concluded that the hypothesis of zero-point radiation energy would have such gravitational effects on the curvature of space that it made the Einstein world impossibly small. During the next two decades the consensus view among experts in quantum mechanics remained a dismissal of Nernst's view. The zero-point radiation energy that formally turned up in quantum field theory was not seen as physically real, but just as a quantity that turned up in calculations and which was unobservable in principle. Without some cut-off in frequency, it would be infinite. Einstein and Pauli shared the majority view that a zero-point energy for blackbody radiation cannot exist.

The only trace I have found of Nernst's idea applied to cosmology, except for Lenz', is a paper by two American physicists who in 1930 proposed a cosmological conjecture that involved the radiation field. "When the electromagnetic field is treated … as an assemblage of independent harmonic oscillators," they wrote, "this leads to the result that there is present in all space an infinite positive energy density." They added the standard view that, "It is infinite because there is supposed to be no upper limit to the frequencies of possible normal modes" [75].

## 5.2. The Cosmological Constant

The cosmological constant is usually traced back to Einstein's 1917 theory of the universe, but in a formal sense it already appeared in his extensive article on general relativity published the year before [76, p. 144]. Einstein briefly considered the field equations with an added lambda term ($\lambda g_{\mu\nu}$), but decided that the extended equations were non-physical. Although a curiosity, it is of some interest that the famous lambda constant originally appeared in a non-cosmological context. It is also worth pointing out that a classical version of the constant can be found more than twenty years earlier [2, p. 109].

In an investigation of the gravitational stability of Newton's infinite stellar universe, the German astronomer Hugo von Seeliger suggested in 1895 that the law of gravitation needed modification at very large distances. At such distances, he argued, a star would be affected by a repulsive force in addition to the attractive gravitational force. Introducing a very small force constant $\lambda$, he wrote the modified law as

$$F(r) = G\frac{m_1 m_2}{r^2}\exp(-\lambda r)$$

The cosmological constant in Einstein's theory was not unlike the constant in Seeliger's earlier work, and it was even designated the same symbol $\lambda$. This was however accidental, for Einstein did not originally know about Seeliger's work. Only later in 1917 did de Sitter make him aware of it.

Einstein's constant can be considered as referring to a vacuum energy density $\rho_{vac}$ associated with $\Lambda$ and a corresponding negative pressure density proportional to the energy density. From the Friedmann equations including a pressure term it follows directly that

$$\rho_{vac} = \frac{\Lambda c^2}{8\pi G} \; ; \; p_{vac} = -\frac{\Lambda c^4}{8\pi G} \; ; \; p_{vac} = -\rho_{vac}c^2$$

The expressions can also be derived from Einstein's field equations. In the parlance of later cosmologists (and with $c = 1$), the equation of state of the cosmological constant is given by the dimensionless parameter $w = p/\rho = -1$.

Einstein was from an early date aware of the connection, but only vaguely and without considering it important. He later commented that in the static world model, "one has to introduce a negative pressure, for which there exists no physical justification … [and] I originally introduced a new member into the field equation instead of the above mentioned pressure" [77, p. 111]. It follows from the cosmological field equations that without a cosmological constant,

$$R^2 = -\frac{1}{\kappa p}$$

In order to be a positive quantity, it requires $p < 0$. If the matter pressure is zero and $\Lambda > 0$, as Einstein originally assumed, the result is instead $R^2 = 1/\Lambda$.

In terms of the critical matter density the vacuum energy can be written as

$$\Omega_{vac} = \frac{\rho_{vac}}{\rho_c} = \frac{\Lambda}{3H^2}$$

When the vacuum expands, the work done to expand it from volume $V$ to $V + dV$ is negative, namely,

$$pdV = -\rho c^2 dV$$

In spite of the expansion, the energy density of the vacuum remains constant (while the energy increases). In general relativity the force of gravity is determined by the combination $\rho + 3p/c^2$. The pressure term can usually be neglected, but in the case of the vacuum we have

$$\rho_{vac} + \frac{3p_{vac}}{c^2} = \rho_{vac} - 3\rho_{vac} = -2\rho_{vac},$$

implying that gravity changes its sign. For this reason the Λ-energy is sometimes described as a form of "anti-gravity."

Contrary to Einstein, Eddington and de Sitter were positively inclined toward the cosmological constant, if for different reasons and to different degrees. To Eddington the cosmological constant was absolutely fundamental, since it provided the yardstick for the radius of closed space. He expressed his commitment as follows: "If ever the theory of relativity falls into disrepute the cosmical constant will be the last stronghold to collapse. To drop the cosmical constant would knock the bottom out of space" [78, p. 104]. With $M$ and $m$ designating the mass of the proton and the electron, respectively, he calculated its value from pure theory:

$$\Lambda = \left(\frac{2GM}{\pi}\right)^2 \left(\frac{mc}{e^2}\right)^4 = 9.8 \times 10^{-55} \text{ cm}^{-2}$$

The basis of Eddington's result was his ambitious and unorthodox theory aiming at unifying cosmology and quantum mechanics. The theory and the theoretical results flowing from it were ignored by the majority of physicists and astronomers [79].

De Sitter knew that the expansion of the universe does not require a positive constant (witness the Einstein-de Sitter model with Λ = 0), but nonetheless seemed to have believed that the cosmological constant was responsible for the expansion of space. Certainly, this was the message of a popular article of 1931, in which he stated that "the expansion depends on the *lambda* alone" [80, p. 9]. Admitting ignorance about the mechanism through which Λ caused the expansion, he chose to see it as an irreducible constant of nature: "To some it may sound unsatisfactory that we are not able to point out the mechanism by which the *lambda* contrives to do it. But there it is, we cannot go beyond the mathematical equations, and … the behavior of *lambda* is not more strange or mysterious than that of the constant of gravitation *kappa*, to say nothing of the quantum-constant *h*, or the velocity of light *c*."

The insight of Λ as a measure of vacuum energy was first explicitly stated by Lemaître in an address of 1933, when he estimated $\rho_{vac} \cong 10^{-27}$ g/cm$^3$ [81; 74, pp. 229-231]. While he offered a physical interpretation of the cosmological constant as a vacuum energy density, he did not connect his interpretation with the zero-point energy of space or otherwise relate it to quantum physics. Lemaître's work attracted no attention at the time and even today it is not well known. He seems not to have considered it important himself. According to the Web of Science database, until February 2014 his paper of 1934 has been cited only 37 times, with 30 of the citations belonging to the period 1999-2013. Between 1949 and 1984 it was not cited at all. The late attention to his work undoubtedly reflects the recent interest in dark energy.

During the first decades after World War II models with a non-zero cosmological constant were held in low esteem. Lemaître continued to advocate a positive constant, but most cosmologists agreed that Einstein had been right in dismissing it as a mistake. The discovery of quasars in 1963 and the discussion of whether or not their high redshifts were of cosmological origin caused some astrophysicists to reconsider models of the kind favored by Lemaître. For example, Nicolai Kardashev suggested a model with Λ = 4.3 × 10$^{-56}$ cm$^2$, ε = 2 × 10$^{-5}$, and an age of the universe of 67 billion years [82].

However, the few models of this type discussed in the late 1960s did not refer to the interpretation of Λ as a vacuum energy density. Moreover, the interest in models with a positive cosmological constant was short-lived, as they turned out not to agree with quasar measurements after all. By 1970 the cosmological constant was out in the cold, once again.

### 5.3. Steps Toward Dark Energy

The question of zero-point energy and related fluctuations in a pure electromagnetic field, or some other quantum field, remained unanswered for a long time. As seen in retrospect, the first evidence – although at first only theoretical – that vacuum energies and fluctuations are indeed real came with Hendrik Casimir's prediction in 1948 of the effect named after him [83; 84]. For a decade or so little interest was paid to Casimir's calculation of an attractive force between two metal plates in a vacuum. Although the first qualitative support from measurements came as early as 1958, it was only in the 1990s that the effect was confirmed quantitatively. Neither Casimir or other researchers did initially think of relating the minute quantum vacuum effect to the vacuum energy associated with the cosmological constant.

A connection of this kind was vaguely perceived by the Russian physicist Erast Gliner, at the Leningrad Physico-Technical Institute, in 1965 and more clearly by his compatriot Yakov Zel'dovich two years later. Spurred by the suggestions that quasar observations might justify models with Λ > 0, in a note of 1967 he stated what more or less tacitly had been known since the 1930s, namely, that Λ corresponds to a vacuum energy and a negative pressure. Zel'dovich's estimate of $\rho_{vac}$ was about 100 times smaller than the figure Lemaître had quoted in 1934.

Only the following year did Zel'dovich seriously consider $\Lambda$ and its consequences, although at the time without referring to the Casimir effect. In an influential review of the cosmological constant problem, Steven Weinberg wrote, obviously with hindsight: "Perhaps surprisingly, it was a long time before particle physicists began seriously to worry about this problem, despite the demonstration in the Casimir effect of the reality of zero-point energies" [85, p. 3].

Arbitrarily assuming a cut-off corresponding to the mass of a proton (~ 1 GeV) Zel'dovich derived a zero-point energy $\rho_{vac}$ of the order $10^{17}$ g/cm$^3$, or $\Lambda \cong 10^{-10}$ cm$^{-2}$, noting that it much exceeded the observational bound on the cosmological constant. This was the beginning of the "cosmological constant problem," namely, that the cosmological constant as calculated from the zero-point energy density of the vacuum $\Lambda_{QFT}$ is hugely larger than bounds imposed by observation. Zel'dovich compared the calculated $\Lambda_{QFT} \cong 10^{-10}$ cm$^{-2}$ with the observational limit $\Lambda_{obs} < 10^{-54}$ cm$^{-2}$ or $\rho_{obs} < 5 \times 10^{-28}$ g/cm$^3$. Although considering the first value to have "nothing in common with reality," he nonetheless suggested that the cosmological constant might arise from the vacuum of quantum field theory [86].

The discovery in 1998 of the accelerating universe driven by a dark energy interpreted as the cosmological constant came as a surprise, but not as a total surprise. At least some cosmologists were well prepared. For example, from considerations of the Hubble diagram and other evidence, in 1975 James Gunn and Beatrice Tinsley suggested that the most plausible cosmological model was a closed, ever-accelerating universe with $\Lambda > 0$ [87]. Referring to Zel'dovich's idea of $\Lambda$ as a vacuum energy density, they thought it made the constant "more acceptable." Ten years later Hans-Joachim Blome and Wolfgang Priester argued that the universe might be in a state dominated by vacuum energy of density of the order $10^{-8}$ erg/cm$^3$ [88].

Finally, the year of 1995 saw the publication of two papers which both argued that the cosmological constant might account for about two thirds of the critical density ($\Omega_\Lambda = 0.6 - 0.7$). Both papers referred to the cosmological constant problem highlighted by Weinberg in particular [85]. According to Jeremiah Ostriker and Paul Steinhardt, "how can we explain the non-zero value of the cosmological constant from a theoretical point of view?" [89]. Lawrence Krauss and Michael Turner referred to the problem in more quantitative terms: "In the context of quantum-field theory there is the fact that a nonzero cosmological constant corresponds to a vacuum energy density, and particle theorists have yet to successfully constrain its value, even to within 50 orders of magnitude of the observational upper limit" [90].

## 6. CONCLUSIONS

The development of modern cosmology differs in certain respects from how other branches of science have developed. In the days of Einstein, Lemaître, and Hubble cosmology did not yet exist as a scientific discipline with its own institutions, university courses, and professional standards. There were physicists, astronomers, and mathematicians who occasionally did cosmological research, but strictly speaking there were no cosmologists. As Bondi admitted in his autobiography, "I always detest being referred to as a cosmologist" [91].

Nor was there any paradigm shared by researchers in the field, such as illustrated by the extended controversy over the steady state theory. The styles and methods of cosmological research differed markedly, ranging between extreme rationalism and extreme empiricism [92]. The situation only changed in the late 1960s with the emergence of a strong consensus theory in the form of the hot big bang model. By following the historical development one not only gains insight in the discoveries and theoretical advances that have shaped modern cosmology, one also gains an understanding of how physical cosmology came into being as a recognized and mature scientific discipline.

This survey gives some background for appreciating the historical foundation of present physical cosmology, but of course it presents a somewhat fragmented picture that leaves out many interesting events and developments. For example, it only deals inadequately with the many unorthodox theories which have not passed the test of history, such as Milne's theory of "kinematic relativity" [47] and Dirac's cosmology of 1937-1938 based on a varying gravitational constant [53, pp. 67-69]. Nor does the essay cover the popular appeal of cosmology, and it also does not discuss the extra-scientific implications of a philosophical, religious and sometimes political nature that for long were integrated elements in the science of the universe and still play some role. These aspects are covered in the historical literature [93].

## ACKNOWLEDGMENTS

I would like to express my thanks to the organizers of this school of special courses for inviting me to give these lectures and also for their hospitality in Rio de Janeiro.